\documentclass[%
aps,twocolumn,floatfix,showpacs,superscriptaddress,amsmath,amssymb]{revtex4-1}

\usepackage{graphicx}
\usepackage{dcolumn}
\usepackage{bm}
\graphicspath{ {./images/} }
\usepackage{color}
\usepackage{hyperref}
\usepackage{footnote}
\usepackage{wrapfig}
\usepackage{subcaption}

\begin{document}

\title{A high-throughput and data-mining approach to predict new rare-earth free permanent magnets}

\author{Alena Vishina}
 \email{alena.vishina@physics.uu.se}
 \altaffiliation{Department of Physics and Astronomy, Uppsala University, Box 516, SE-75120, Uppsala, Sweden}
\author{Olga Yu. Vekilova}
\affiliation{Department of Physics and Astronomy, Uppsala University, Box 516, SE-75120, Uppsala, Sweden}
\author{Torbj\"orn Bj\"orkman}
\affiliation{Department of Natural Sciences, $\mathrm{\mathring{A}}$bo Akademi, FI-20500 Turku, Finland}
\author{Anders Bergman}
\affiliation{Department of Physics and Astronomy, Uppsala University, Box 516, SE-75120, Uppsala, Sweden}
\author{Heike C. Herper}
\affiliation{Department of Physics and Astronomy, Uppsala University, Box 516, SE-75120, Uppsala, Sweden}
\author{Olle Eriksson}
\affiliation{Department of Physics and Astronomy, Uppsala University, Box 516, SE-75120, Uppsala, Sweden}
\affiliation{School of Science and Technology, Örebro University, SE-701 82 Örebro, Sweden}

\date{\today}

\begin{abstract}
We present an application of a high-throughput search of new rare-earth free permanent magnets focusing on  3d-5d transition metal compounds. The search involved a part of the ICSD database (international crystallographic structural database), together with tailored search criteria and electronic structure calculations of magnetic properties. Our results suggest that it possible to find candidates for rare-earth free permanent magnets using a data-mining/data-filtering approach. The most promising candidates identified here are Pt$_2$FeNi, Pt$_2$FeCu, and W$_2$FeB$_2$. We suggest these materials to be a good platform for further investigations in the search of novel rare-earth free permanent magnets.
\end{abstract}

\maketitle

\section{Introduction}

High performance permanent magnets, or materials with large energy product, are needed for a large number of applications, such as electric vehicle motors and generators, wind mills, loud speakers, and relays. The materials properties needed for these application, many of which represents a `green' technology of energy conversion, are an essentially high saturation magnetization, a large Curie temperature and a large magnetic anisotropy. Most permanent magnets that are used today are either ferrites 
\cite{craik} or rare-earth (RE) containing compounds, such as Nd$_2$Fe$_{14}$B \cite{coey}. In fact, most high performance magnets contain RE materials such as Pr, Nd, Sm, Tb, or Dy \cite{perm_mag}, which makes them expensive, while some of the RE elements (like Dy) are rapidly decreasing in availability and therefore bear an economic risk. With the increase in the number of electric vehicles and the usage of wind power generators, there is a growing need to reduce or eliminate the usage of RE magnets by finding rare-earth free alternatives, which at the same time are expected to at least show the same prize-performance. Due to the high interest in the field, various materials \cite{mat2,mat4,mat8}, nanostructures \cite{mat1,mat3,mat5,mat7}, thin films \cite{film1,mat6}, and new compounds have been proposed. In addition, new 
techniques, such as machine learning \cite{mach-learn}, have been suggested in the quest for finding RE free permanent magnets. Most of the papers devoted to this effort describe the experimental investigations of a single compound, or a smaller 
group of materials with similar structure \cite{mat2,mat4,mat9,mat10,mat11,mat12,mat13,mat14,mat15}.

Experimental studies of the magnetic properties of a single material requires first synthesizing it and performing a number of measurements in order to characterize the required quantities; e.g. crystal structure, saturation- and remanent magnetic moment, coercive field, and ordering temperature. This is a cumbersome and time-consuming effort, that makes a systematic study of a large number of materials, not only expensive, but also extremely lengthy. Even though there exist experimental high-throughput studies, a detailed analysis of materials properties as described above can be extremely time consuming. Theoretical calculations, especially based on density functional theory (DFT), have developed at a rapid speed and are now found to reproduce essentially all magnetic properties that are needed for a suitable permanent magnet \cite{ASD}. 
In fact there are even a few cases when theory has made predictions of materials with suitable characteristics as RE free permanent magnets. For instance, Fe-Co alloys that were tuned to a specific concentration and tetragonal strain were predicted to have a large magnetic moment and magnetic anisotropy \cite{mat9}, and subsequent experimental work \cite{mat10} confirmed this suggestion. 
It should be noted here that the most challenging aspect of these studies is the calculation of the magnetic anisotropy energy (MAE), simply because its value is tiny. However, since the pioneering works of Brooks \cite{Brooks}
several attempts \cite{Fletcher1954, Slonc, Kond, Bruno,Daalderop} have shown that, with few exceptions, theory reproduces the easy axis orientation and the right order of magnitude of the MAE. Hence, theory is a good alternative that allows to identify new phases and compounds that are suitable candidates for RE free permanents. We have therefore combined density functional theory, a high-throughput search from the International Crystallographic Structural Database (ICSD) \cite{ICSD}, and a data-filtering through a large amount of data to identify new RE free materials that can be proposed as new permanent magnets.

This approach has recently been applied for the crystal structure predictions \cite{cr1,cr2},  in the search for materials with specific properties \cite{melt,thermo,bond}, such as new two-dimensional materials \cite{2d,2d-elec}, new perovskites \cite{pero}, possible high-temperature superconductors \cite{htc,hts2}, new scintillator materials \cite{sci}, topological insulators \cite{top}, spin-gapless semiconductors \cite{spin_gap}, and materials design in general \cite{design, Eriksson2018}. From our investigations we have identified three compounds that fulfill the criteria of being suitable as permanent magnets, without containing RE elements.  

\section{Computational methods}\label{sec:method}

Initial screening was performed using a database of ~10 000 calculations derived from compounds in the ICSD database containing the elements Cr, Mn, Fe, Co and Ni, and which have no more than 16 atoms in their primitive unit cell. These calculations were performed using the full-potential linear muffin-tin orbital method (FP-LMTO), including spin-orbit interaction as implemented in the RSPt code \cite{rspt1, rspt2}, with the AM05 functional \cite{am05} for exchange and correlation. Calculation time depended strongly on a system and varied between several and several hundreds CPU hours.

For determining the magnetic ground state of materials, i.e. ferromagnetic (FM)  or antiferromagnetic (AFM), Vienna Ab Initio Simulation Package (VASP) \cite{vasp1,vasp2,vasp3,vasp4} was used within the Projector Augmented Wave (PAW) method \cite{PAW}, along with the Generalized Gradient Approximation (GGA) in Perdew, Burke, and Ernzerhof (PBE) form \cite{PBE}.

For the phase stability check of the materials the differences in enthalpies were calculated between the systems and proper stable binary compounds chosen from the corresponding phase diagrams. For instance the stability of Pt$_2$FeNi is checked by calculating the formation enthalpy in the form 
\begin{equation*}
\Delta H = H_{FeNiPt_2}-H_{FePt}-H_{NiPt},
\end{equation*}
where $H_{FePt}$ and $H_{NiPt}$ are the enthalpies of FePt and NiPt, respectively.

 The MAE calculations were done with the the RSPt code, as $\Delta E = E^{pl} - E^{c}$, where $E^{c}$ and $E^{pl}$ are the total energies with the magnetization directed along and perpendicular to the $c$-axis. The results were obtained with the tetrahedron method with Blöchl correction for the Brillouin zone integration \cite{Blochl}; the converged k-point Monkhorst-Pack meshes \cite{MP} between $28\times28\times28$ and $34\times34\times34$ were used for the calculations depending on a material. Materials with the positive $\Delta E$ have uniaxial magnetic anisotropy required for a permanent magnet. Calculation of MAE depending on the material required CPU time between several hundreds and several thousands hours.
 
 Magnetic hardness parameter was calculated as  $\kappa = \sqrt{\Delta{E}/\mu_0 M_S^2}$ \cite{hardness}, where $M_S$ is saturation magnetization and $\mu_0$ is the vacuum permeability.
 
 Curie temperature was estimated using Monte Carlo simulations implemented within Uppsala  atomistic  spin  dynamics (UppASD) software \cite{ASD} starting from the magnetic parameters computed with RSPt code. The ASD simulations were performed on a $35\times 35\times 35$ supercell, with periodic boundary conditions, using the  exchange parameters calculated with RSPt code within the first nine coordination shells.
 
 Sumo package \cite{sumo} was used for the density of states (DOS) plots. 
 
 \section{High-throughput and Screening}

There are several criteria which are necessary for a material to be a strong permanent magnet. Among these are ferromagnetic ordering, high saturation magnetization ($M_S> 1$ T), high Curie temperature (T$_c >$ 400$^{\circ}$C), and, more importantly, large uniaxial magnetic anisotropy energy (MAE) ($> 1$ MJ/m$^3$). Even though a large magnetization is desirable for permanent magnets the ratio between magnetization and MAE is even more important since it determines the hardness $\kappa$ of a magnet and $\kappa \ge 1$ is needed for a hard magnet in order to resist the self-demagnetization during its fabrication \cite{coey}.

Evaluating the saturation magnetization from first principles theory is rather straight-forward, and can be done for a large number of materials in a relatively efficient manner. The calculation of MAE, on the other hand, is rather complex and involves tiny energy differences of the order of $\mu$eV \cite{Daalderop,Trygg}. Magnetic anisotropy is the result of relativistic effects, primarily the spin-orbit coupling, combined with the exchange splitting of the energy bands, and hence requires first-principles calculations using a relativistic formulation of the Kohn-Sham equations.

\begin{figure}[h!]
 \caption{(Color online) The steps of high-throughput search and screening with the number of structures left after each step.}
 \centering
 \includegraphics[scale=0.1]{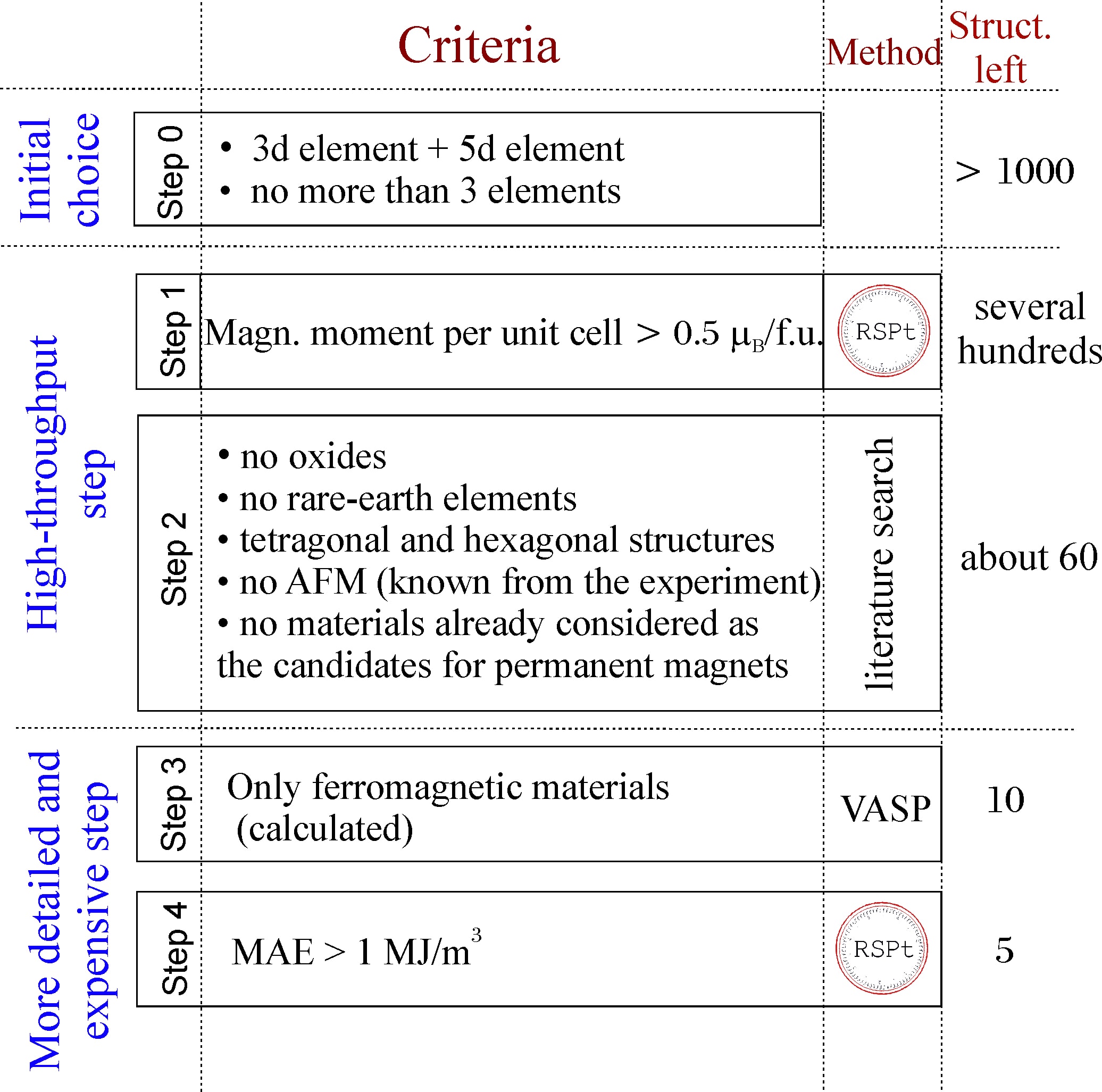}
 \label{Diagram}
\end{figure}

The search presented here was performed in two parts (see Figure \ref{Diagram}): first, by performing a high-throughput screening of a very large number of possible compounds a limited list of materials was identified, and then the more detailed calculations were used for the second step. The workflow of this approach was such that initially a high-throughput screening was made, using data sets provided by  the  ICSD \cite{ICSD} database, for which we applied a set of initial screening parameters. In the present study we report on results of a search among compounds that have at least one  3d- and one 5d-element. Such materials have the potential to have a significant MAE, since the 3d element would provide significant exchange interaction and magnetic ordering temperature, while the 5d element would guarantee large spin-orbit coupling that is needed for the MAE. This initial choice left us with a list of more than a thousand systems. Since it is known that spin-orbit effects influence the MAE most effectively for non-cubic materials, we also screened to identify tetragonal and hexagonal materials. These two initial screening steps did not rely on any theoretical calculations, only information from the ICSD \cite{ICSD} database was used. 

First principles calculations (see the \emph{Computational methods} section, Sec.\ref{sec:method}) of the materials identified in the first two screening steps was then made, and further screening step was introduced, where we considered only materials with calculated magnetic moment higher than 0.5 $\mu_B$/f.u. Among these compounds several oxide materials were identified. Since transition metal oxides often have a correlated electronic structure, in such a way that single determinant theories such as density functional theory fail, we excluded these compounds. It should be emphasized that these oxides were excluded not because they were considered uninteresting for applications as permanent magnets, but because the theory employed here is known to have questionable accuracy when it comes to describing magnetic properties, something which in particular applies to the MAE.

To further reduce the number of compounds that were subjected to careful, high precision calculations of the saturation moment and MAE, we excluded materials consisting of more than three elements.  All these screening steps left us with less than 60 possible candidates. Some, but not all, of these compounds have previously been investigated with regard to their magnetic properties, at least the magnetic structure has been reported for some of them. All the materials that are known in the literature to be antiferromagnetic or paramagnetic (PM) at room temperature were then excluded from the list, which reduced the number of potential compounds to 25. These systems  were selected for high-precision calculations of the magnetic anisotropy and the ordering temperature, giving information about the four most important, intrinsic magnetic properties of potential new permanent magnets; ferromagnetic ordering, large saturation moment, significant ordering temperature and a relatively large magnetic anisotropy. 
 
\section{Results}
\subsection{Findings from  high-throughput calculations}
We applied the screening steps outlined above to materials that contain necessarily one 3d and one 5d element, while at the same time allowing the material to have no more than three elements in total. All the crystallographic data were taken from the ICSD \cite{ICSD} database. 
The screening steps employed here allowed to single out a number of materials that have already been suggested and investigated as systems with high magnetic anisotropy in previous works \cite{MnAu,CoPt_exp1,CoPt_th1,CoPt_exp2,CoPt_th2,CoPt_th3,FePt_exp,Fe2Ta_Fe2W}. In Table \ref{table:1} we list the materials properties of these compounds. Note that several of the compounds in Table \ref{table:1} have previously been recognized from theoretical and/or experimental work to have promising materials properties for applications as permanent magnets, e.g. FePt. Hence the fact that the screening method employed here identifies these systems, renders credibility to its power as a predictive tool. We note here that of the compounds listed in Table \ref{table:1}, Fe$_2$W has previously been investigated only by theoretical means. The Curie temperature T$_c$ was calculated for this material since is has not been reported before.

Unfortunately, none of the materials listed in Table \ref{table:1} is suitable for applications that need large mass/volume, since they involve costly elements. The exception to this is Fe$_2$W, but for this compound the saturation moment is too low to make it interesting for applications. However, alloying of this compound may improve the saturation moment. Such an investigation is outside the scope of the present work. 

\begin{table*}[t]
\caption{Well known materials with appropriate magnetic characteristics for permanent magnet applications, that came out of the screening steps proposed in this investigation. Listed are the ICSD database number, space group, MAE, magnetic moment ($\mu$) per 3d atom, and Curie temperature (T$_c$). In parenthesis we note measured (exp) or previously calculated (th) data. References to previous work are also included in parenthesis. T$_c$ for Fe$_2$W was calculated in the present investigation.}
\begin{tabular}{c c c c c c c c} 
 \hline \hline
 Material & ICSD & Space & MAE (exp) & MAE (th) & $\mu$ & T$_c$ \\ 
  & number & group & MJ/m$^3$ & MJ/m$^3$ & $\mu_B$/3d  & K \\
 \hline
 Au$_4$Mn & 657182 & 87 & 0.3 \cite{MnAu} &   & 4.3 \cite{MnAu} & 370 \cite{MnAu} \\ 
 CoPt & 197572 & 123 & 4.0 \cite{CoPt_exp1} & 6.6 \cite{CoPt_th1} & 1.76 \cite{CoPt_exp2} & 853 \cite{CoPt_exp1} \\
  & & & 3.0 \cite{CoPt_exp2} & 6.9 \cite{CoPt_th2} & & \\
  & & & 4.5 \cite{CoPt_th3} & & \\
 FePt & 633191 & 123 & 10.0 \cite{FePt_exp} & 11.0 \cite{CoPt_th1} & 2.8 \cite{FePt_exp2} & 773 \cite{FePt_exp} \\
 Fe$_2$W & 634058 & 194 &  & 0.87 \cite{Fe2Ta_Fe2W} & 0.43 \cite{Fe2Ta_Fe2W} & 320 \\
 \hline \hline
\end{tabular}
\label{table:1}
\end{table*}

In addition to the known structures listed in Table \ref{table:1}, a set of hitherto unexplored compounds was identified, which involved about twenty materials. For these compounds, more detailed and accordingly time-consuming considerations were made, where in particular focus was paid to a large magnetocrystalline anisotropy with uniaxial orientation, ferromagnetic ordering, high saturation magnetization, and  large exchange interaction. The latter guarantees a sufficiently high Curie temperature. Of these compounds, several 
failed to match one or more of the above mentioned key requirements for permanent magnets. The calculations showed that some of them have an antiferromagnetic ground state. Examples of such systems are W$_2$MnB$_2$, NiMnAs, GeMnTa, and Co$_2$Ta (see Appendix for the full list of materials found to be AFM).  Materials that were found to  have a planar easy axis ($\Delta E < 0$) or a MAE that is not sufficiently high, were also excluded from the final list of proposed rare-earth free permanent magnets. Examples of systems that were put aside are Al$_8$Fe$_4$Hf and Fe$_7$W$_6$. Although these materials are not identified here as having suitable magnetic properties for permanent magnet applications, we note that alloying of these systems, or investigations of the temperature dependence of the MAE, may render them suitable. However, such studies are outside the scope of the present investigation. 

\begin{figure}[h]
\includegraphics[scale=0.28]{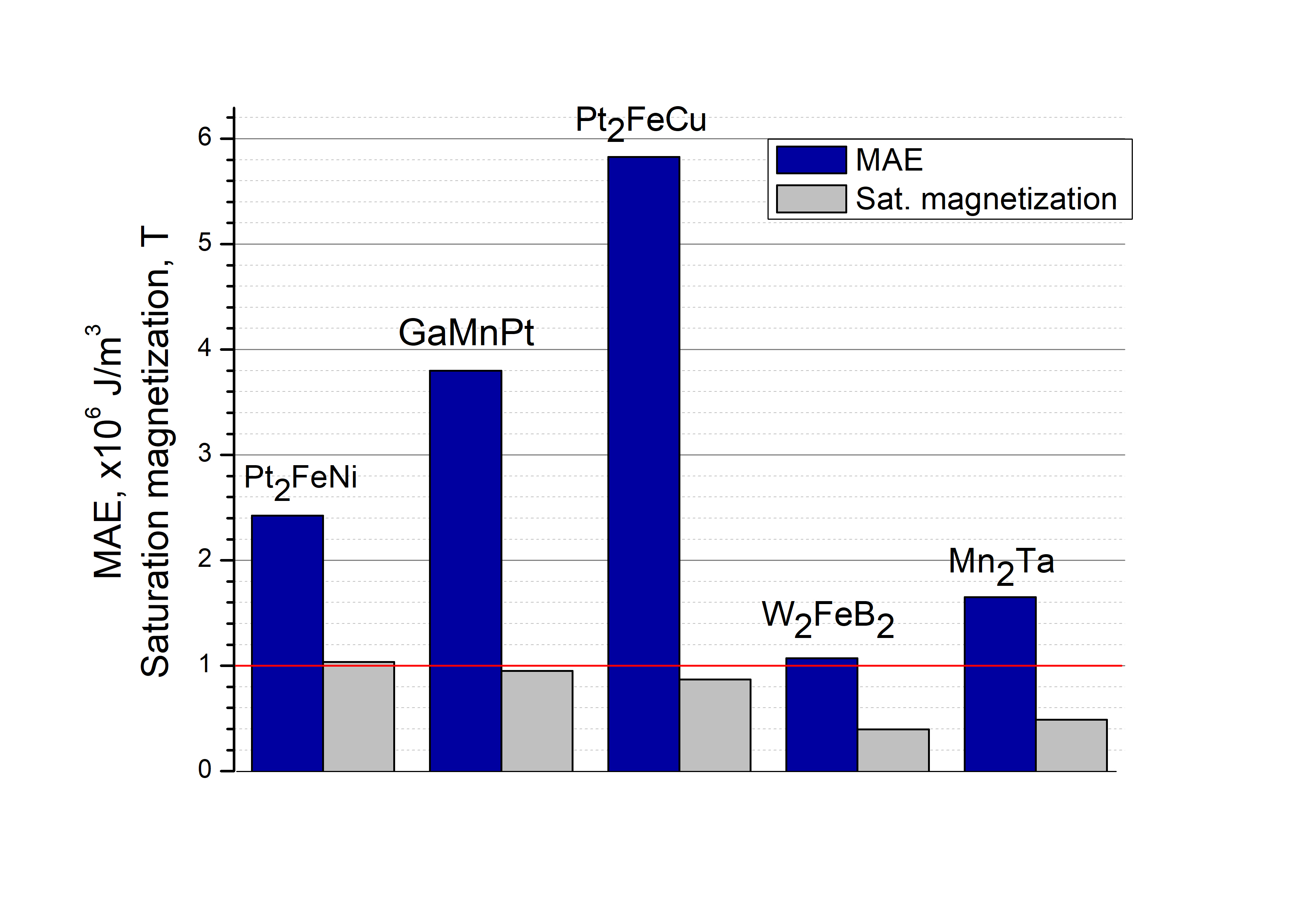}
\caption[justification=justified]{(Color online) MAE and saturation magnetization for several compounds, that have not been investigated for magnetocrystalline anisotropy before (apart from Pt$_2$FeNi that which was studied in the form of L1$_0$ alloy films \cite{FeNiPt}). The red line shows MAE = 1 MJ/m$^3$ and $M_s=1$ T that are considered the necessary criteria of a strong permanent magnet.} 

\label{MAEfig}
\end{figure}

Apart from the previously known materials with good magnetic properties for permanent magnet applications, listed above, five additional promising compounds were identified, see Fig.~\ref{MAEfig}. To our knowledge the magnetic properties of these materials have not been investigated in details before. These systems seem to possess magnetic properties that are required to be considered as a candidate phase for permanent magnets. The calculated magnetocrystalline anisotropy is uniaxial and well above the threshold of 1 MJ/m$^3$. For three of them also the saturation magnetization is larger than 1~T, only Mn$_2$Ta and W$_2$FeB$_2$ fall behind with magnetization values around 0.5~T, see Fig.~\ref{MAEfig}.  It may be noted here that Pt$_2$FeCu has an extremely large MAE of about 6 MJ/m$^3$. In fact few materials can compete with such a large uniaxial anisotropy. Only materials such as FePt and CoPt provide MAE values of the same order of magnitude, see Tab.~\ref{table:1}. The MAE of GaMnPt is also very significant and is interesting from a practical point of view, since it has even lower Pt concentration than the tetragonal Pt$_2$FeCu system. 

\begin{figure}[h]

\begin{subfigure}{0.5\textwidth}
\includegraphics[scale=0.25]{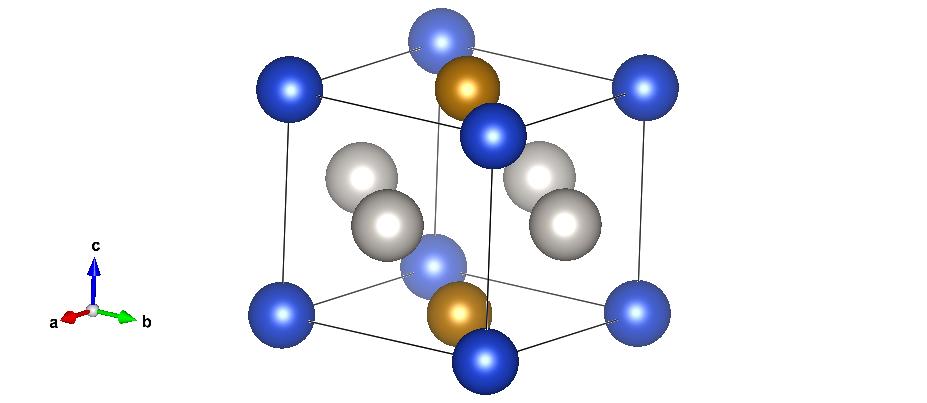} 
\label{Pt2FeCu_cell}
\end{subfigure}
\begin{subfigure}{0.5\textwidth}
\includegraphics[scale=0.2]{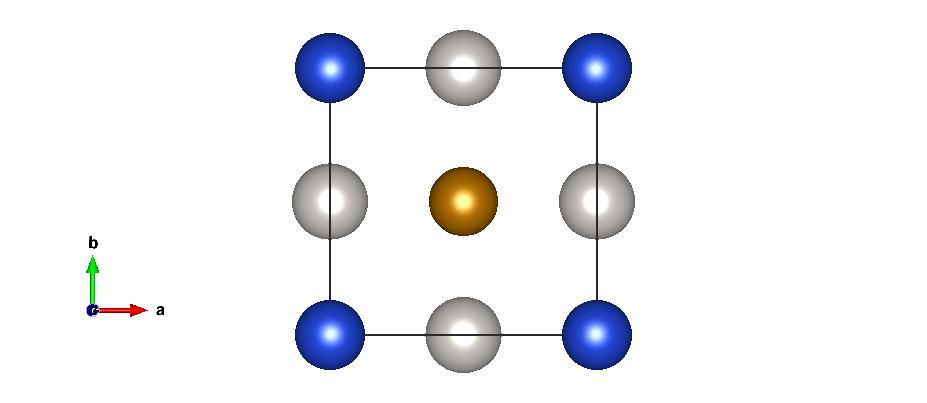}
\label{Pt2FeCu_top}
\end{subfigure}
 
\caption{(Color online) Tetragonal unit cell of Pt$_2$FeM (M = Cu, Ni). Iron atoms are shown with the brown spheres, M atoms are blue, and Pt atoms are grey.}
\label{Pt2FeM}
\end{figure}

\begin{figure}[h]

\begin{subfigure}{0.5\textwidth}
\includegraphics[scale=0.25]{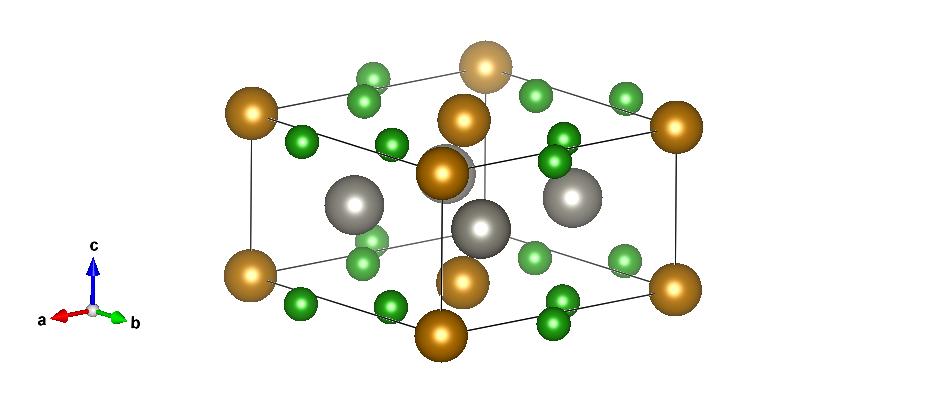} 
\label{W2FeB2_cell}
\end{subfigure}
\begin{subfigure}{0.5\textwidth}
\includegraphics[scale=0.2]{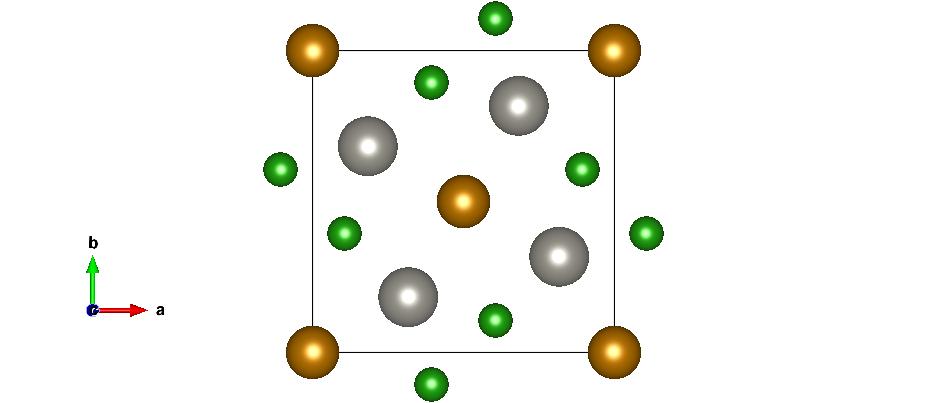}
\label{W2FeB2_top}
\end{subfigure}
 
\caption{(Color online) Tetragonal unit cell of W$_2$FeB$_2$. Iron atoms are shown with the brown spheres, B atoms are green, and W atoms are grey.}
\label{W2FeB2}
\end{figure}

Taking a closer look at the systems in Fig.2, by investigating the magnetic exchange parameters and magnetic coupling, revealed that GaMnPt an Mn$_2$Ta have a more complex magnetic order and are strictly speaking not ferromagnets.
Despite GaMnPt being found to be ferromagnetic at finite temperatures by several experimental groups \cite{GaMnP2,GaMnP,GaMnPt3,GaMnPt4} we found that in the ab-initio calculations it shows deviations from the collinear configuration. We believe that the observed ferromagnetic state of GaMnPt is possible being a result of careful and tailored alloying, investigation of lattice defects, or just as a finite temperature effect. However, to investigate this further is outside of the scope of this paper.
Therefore, these two compounds are excluded from the suggested list of new materials, and their magnetic properties will be discussed elsewhere. For the remaining three tetragonal phases all calculated magnetic properties are summarized in Table~\ref{table:2}.  The unit cells of these materials can be seen in Fig.~\ref{Pt2FeM} and Fig.~\ref{W2FeB2}. Pt$_2$FeCu and Pt$_2$FeNi have the same crystal structure, with the same site occupied either by Cu or by Ni. Additional structural information can be found in Appendix B.  

One of the materials listed in Table \ref{table:2}, Pt$_2$FeNi, was previously studied in the form of L1$_0$ alloy films \cite{FeNiPt,FeNiPt2}). This system was found in experiments to have a uniaxial magnetic anisotropy of 0.9 MJ/m$^3$, which qualitatively agrees with the theoretical data presented here of bulk  Pt$_2$FeNi (see Table \ref{table:2}). To address the stability of  Pt$_2$FeNi as a bulk compound we analyzed theoretically the phase stability. Its formation enthalpy was found to be $\Delta H = -0.19 $ eV/f.u., which indicates the bulk crystalline phase to be stable. 
From the three materials in Tab.~\ref{table:2} Pt$_2$FeNi has the best average performance with a MAE above 2 MJ/m$^3$ and a saturation magnetization of 1~T. Its Curie temperature is still below room temperature but it might be increased e.g. by alloying or doping.  

\begin{figure}[h]
\centering
\includegraphics[scale=0.06]{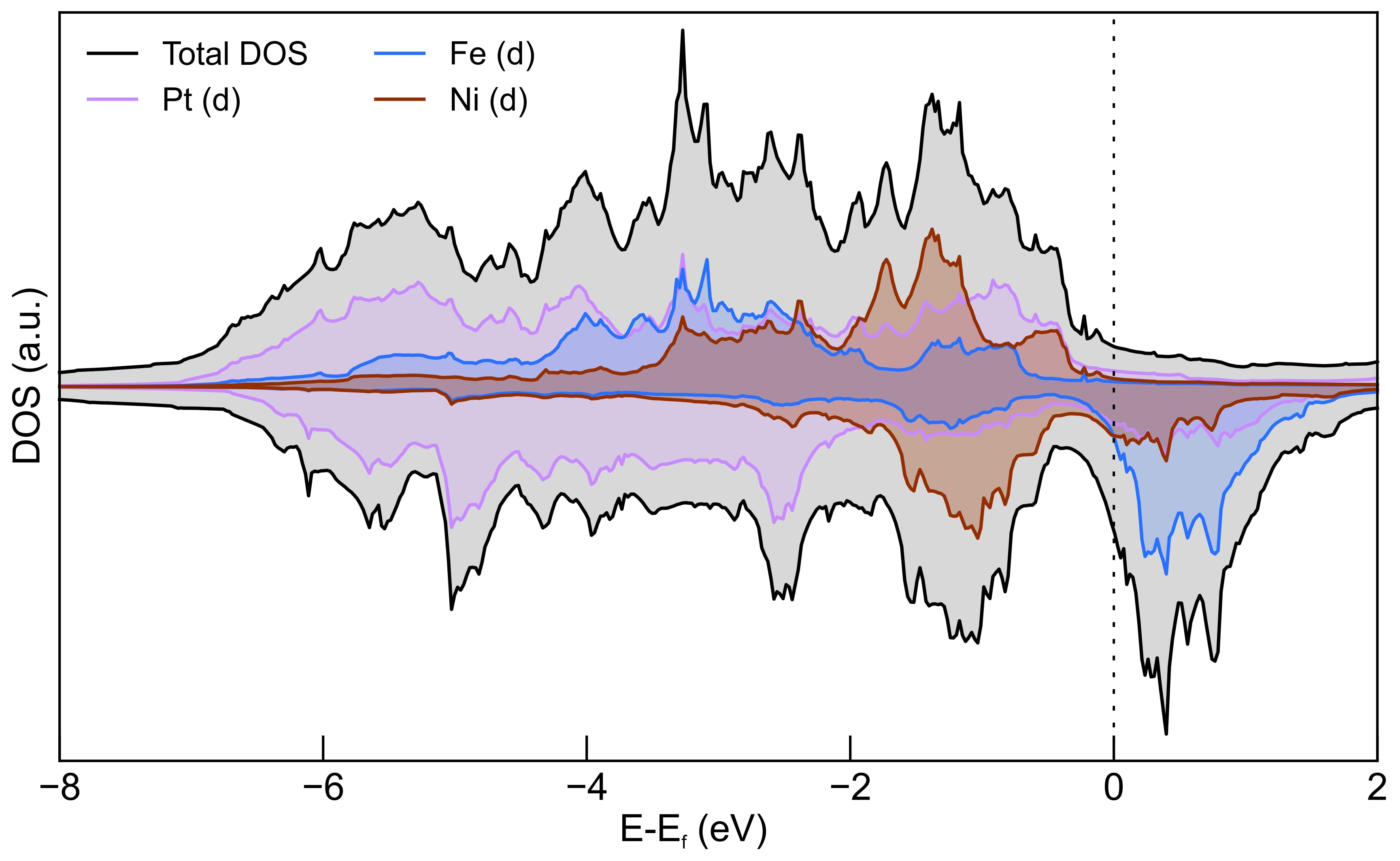}\\
\includegraphics[scale=0.06]{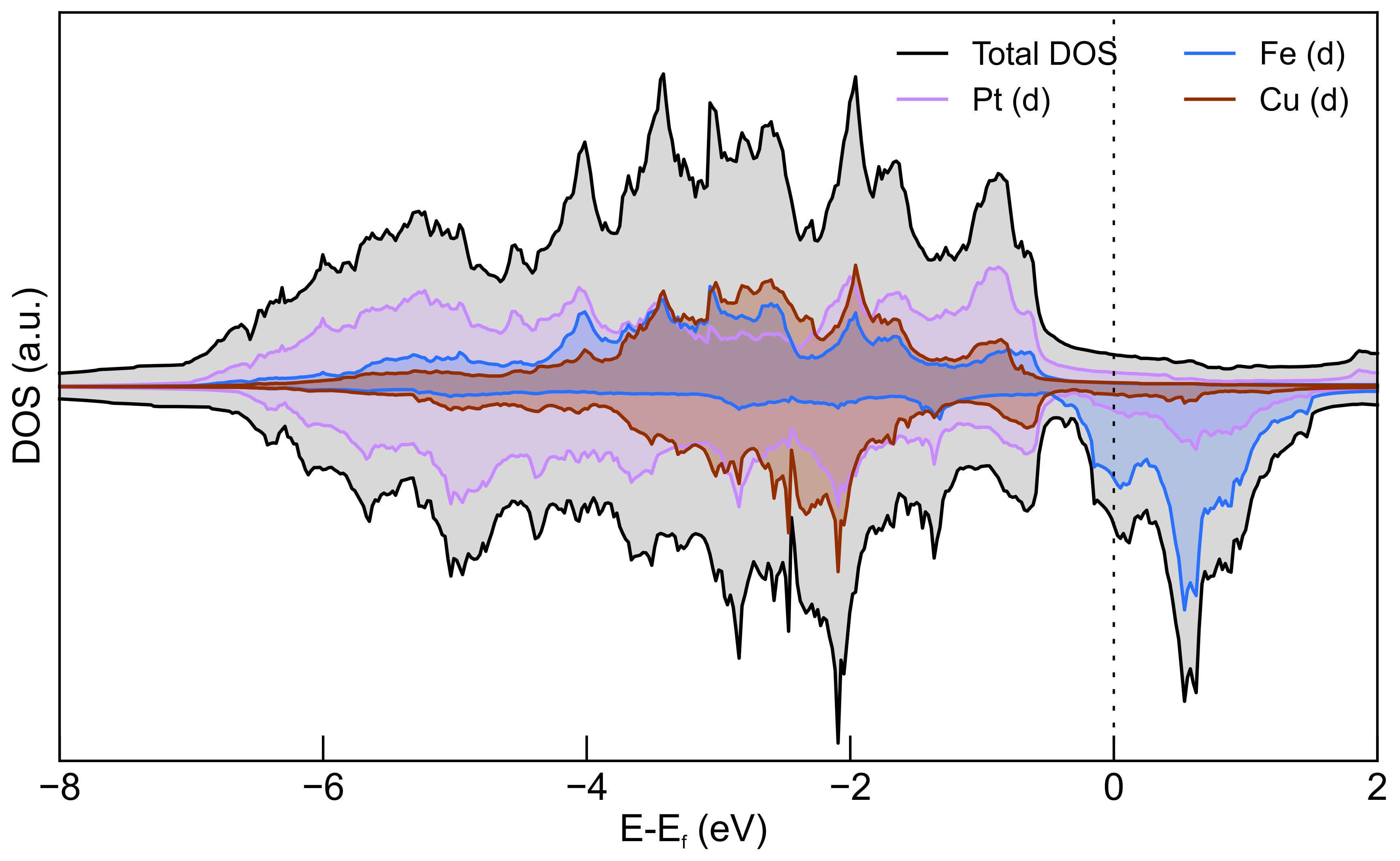}
\caption{(Color online) Calculated spin polarized density of states for the experimental crystal structures of Pt$_2$FeNi(top) and Pt$_2$FeCu (bottom). Only the d-states are shown with spin-up density of states at the top and spin-down at the bottom.}
\label{Pt2FeM_DOS}
\end{figure}

\begin{figure}[h]
 \centering
 \includegraphics[scale=0.06]{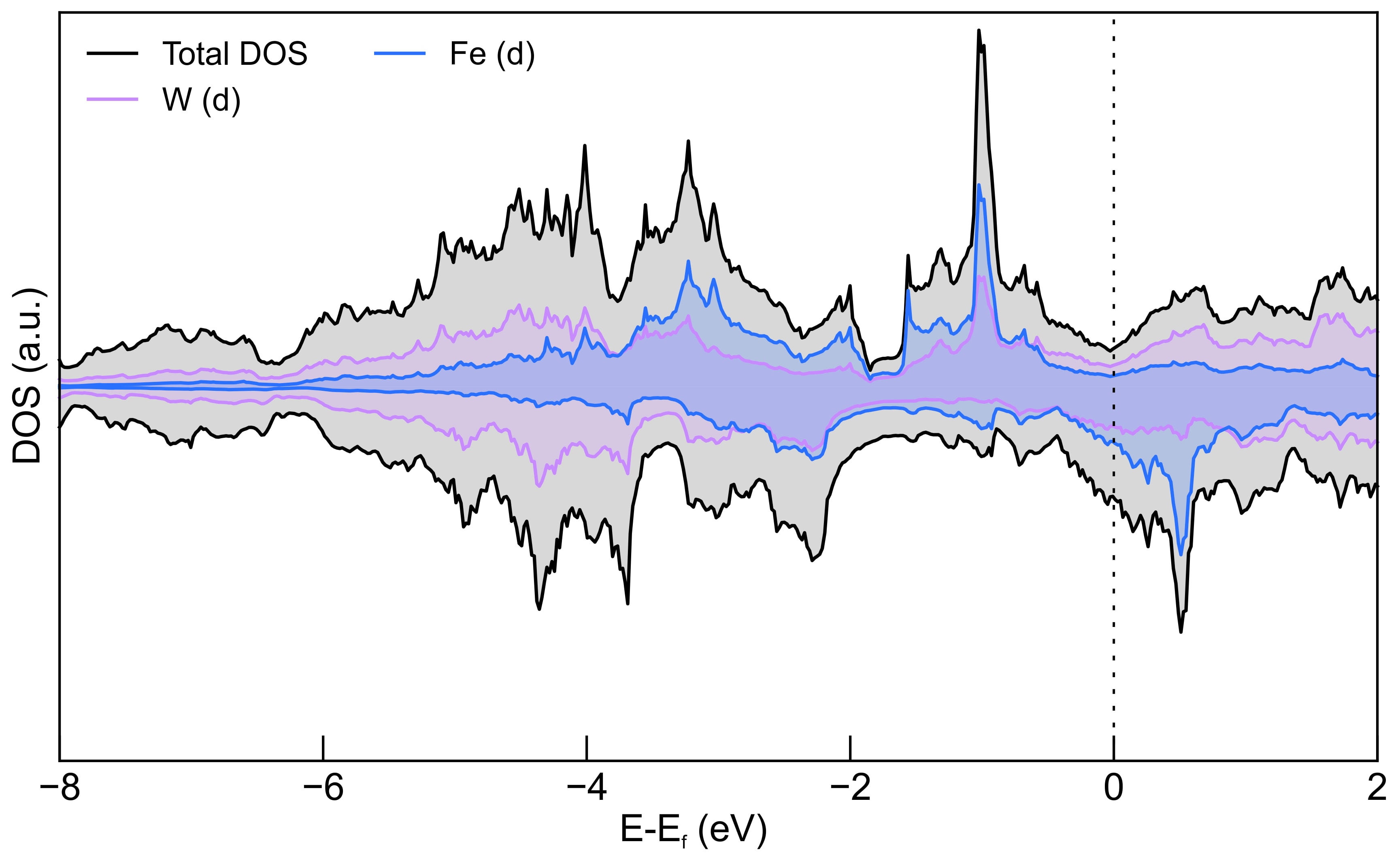}
\caption{(Color online) Calculated spin polarized density of states for the experimental crystal structure of W$_2$FeB$_2$, spin-up (top) and spin-down (bottom). Only the d-states are shown with spin-up density of states at the top and spin-down at the bottom.}
\label{W2FeB2_dos}
\end{figure}

Density of states for the three materials listed in Table II, are plotted in Fig.~\ref{Pt2FeM_DOS} and Fig.~\ref{W2FeB2_dos}. Pt$_2$FeCu and Pt$_2$FeNi, which have similar magnetic moments, have the similar DOS, with the spin-up channel almost completely filled and the spin-down channel partially occupied. In these compounds it is possible to discern from the partial DOS, hybridization between 3d orbitals of Fe and Ni(Cu) and 5d orbitals of Pt. For W$_2$FeB$_2$ hybridization between 3d and 5d orbitals is also obvious. However, for this compound the exchange splitting is not as marked as for Pt$_2$FeCu and Pt$_2$FeNi, and both spin-channels are partially filled. 

From Table~\ref{table:2} it becomes clear that two of the three compounds are not suitable  for applications that require larger volume or mass, since they  contain Pt, which is expensive. In fact, from a prize performance view point, of the materials listed in Table \ref{table:2}, W$_2$FeB$_2$ has the highest potential for bulk applications in large volume or mass, although one would have to find ways to increase the ordering temperature and preferably also the saturation moment.

\begin{table}[h!]
\centering
\caption{Calculated MAE, saturation magnetization, Curie temperature, and magnetic hardness parameter for materials that previously have not been investigated with respect to the magnetocrystalline anisotropy (apart from Pt$_2$FeNi that which was studied in the form of L1$_0$ alloy films \cite{FeNiPt}). }
\begin{tabular}{c c c c c c c} 
 \hline \hline
 Material & ICSD & Space & MAE (th) & Sat. magn. & T$_c$  & $\kappa$ \\ 
  & number & group &  MJ/m$^3$ & T & K & \\
 \hline
Pt$_2$FeNi & 42564 & 123 & 2.42 & 1.04 & 230 & 1.68\\ 
Pt$_2$FeCu & 53259 & 123 & 5.83 & 0.87 & 30 & 3.11 \\
W$_2$FeB$_2$ & 43016 & 127 & 1.07 & 0.40 & 165 & 2.90 \\
 \hline \hline
\end{tabular}
\label{table:2}
\end{table}
 
\subsection{Post high-throughput refinement}
Building on the high throughput study presented in the previous section the outcome can be used to further improve the magnetic properties by careful analysis and modification of the existing phases. Here, we focus on the tetragonal Pt$_2$MM$^{'}$ phase. 
One may speculate that isoelectronic versions of them would also be good candidates as permanent magnets. Typically the chemistry of isoelectronic elements is similar, and for this reason a substitution of isoelectronic elements may result in similar phase and structural stability. One may speculate that alloying of Ni and Pd for Pt in Pt$_2$FeNi and Pt$_2$FeCu may give materials with desired magnetic properties for permanent magnets. In particular, Ni substitution is desired in order to reduce the cost of the compound. 

At the same time one may try to resolve the issue of the low Curie temperature, that we obtained for Pt$_2$FeNi and Pt$_2$FeCu (which otherwise have sufficiently high MAE and saturation magnetization) by replacing either of 3d metals with a different element. For that reason we decided to look into similar compounds that contain cobalt, which usually increases the T$_c$. Pt$_2$CoNi and Pt$_2$FeCo are not in the current version of ICSD that was used here. However they have been synthesized in the form of nanoparticles \cite{FeNiCo}, which suggests that bulk versions of these systems might be possible. We relaxed these materials in the same space group as Pt$_2$FeNi and Pt$_2$FeCu; and found them to be ferromagnetic. Saturation magnetization, MAE, and other properties of the Co-based materials can be found in Table \ref{table:3}. With the Curie temperature increased by the addition of cobalt above the value that is required for a good permanent magnet,  Pt$_2$CoNi and Pt$_2$FeCo represent the best set of properties out of the materials we found after a high-throughput search and screening.
\begin{table}[h!]
\centering
\caption{Calculated MAE, saturation magnetization, Curie temperature, and magnetic hardness parameter for the materials proposed based on the findings of high-throughput run.}
\begin{tabular}{c c c c c c c c} 
 \hline \hline
 Material  & Space & Volume/ion & MAE & Sat. magn. & T$_c$  & $\kappa$ \\ 
  & group & $\mathrm{\mathring{A}}^3$ &  MJ/m$^3$ & T & K & & \\
 \hline
Pt$_2$CoNi & 123 & 13.47 & 5.46 & 0.83 & 385 & 3.16 \\ 
Pt$_2$FeCo & 123 & 13.75 & 6.83 & 1.24 & 605 & 2.36 \\
 \hline \hline
\end{tabular}
\label{table:3}
\end{table}

\section{Discussion and conclusions}

We propose using a materials specific screening method combined with a high-throughput approach of the electronic structure, for the search of possible candidates for rare-earth free permanent magnets.
In order to test the method we have performed a data-mining search among the known crystal structures of the ICSD database \cite{ICSD} containing 3d and 5d elements. A number of materials have been found, in particular three promising compounds, which have previously not been studied with respect to the magnetic anisotropy. The three systems identified have magnetic properties that satisfy the requirements of having a high saturation magnetization and a large MAE, necessary for a material to be used as a high-performance permanent magnet. The identified compounds are: Pt$_2$FeNi, Pt$_2$FeCu, and W$_2$FeB$_2$. GaMnPt and Mn$_2$Ta were found to have a complex magnetic ground state which is not desirable for permanent magnet applications but might have potential for other magnetic applications.   

From MAE point of view the Pt$_2$FeCu system is the most favorable but as can be seen from Table\,\ref{table:2} the Curie temperature is very low (30 K) whereas the $T_C$ of the sister compound Pt$_2$FeNi is 230 K. Thus replacing Cu by a magnetic element improves the transition temperature. We investigated the sister compounds Pt$_2$CoNi and Pt$_2$FeCo which, although not yet in the ICSD database, have been synthesized in the form of nanoparticles \cite{FeNiCo}. Replacing some of the elements by cobalt indeed increased the T$_c$ to 385 K for bulk Pt$_2$CoNi and 605 K for Pt$_2$FeCo, while keeping the high MAE and saturation magnetization. Hence  Pt$_2$CoNi and Pt$_2$FeCo posses the best set of properties out of the materials we found after a high-throughput search and screening.

The clear drawback is the high Pt concentration. However, these systems may be relevant for applications that do not rely on large samples, e.g. as write heads.
Furthermore, as discussed in the previous section isoeletronic partial replacement of Pt might be a solution but is still under consideration and beyond the scope of the present paper.
 Out of the five listed compounds, the most interesting one for applications using larger masses or volumes, seems to be W$_2$FeB$_2$ which has a high MAE $ = 1.07 $ MJ/m$^3$, although its saturation magnetization is slightly too low to meet the requirement of a good permanent magnet, it is found to be 0.40 T. Further alloying to increase the saturation moment and ordering temperature, would make this system even more competitive from an application point of view.  

Furthermore, Mo and Cr substitution for W in W$_2$FeB$_2$ could be an interesting avenue to search for rare-earth free permanent magnets. Since the calculated ordering temperature of this compound is only 165 K (Table 2), the Cr substitution (which is an element that traditionally has a larger interatomic exchange interaction than W), may be a way to increase the T$_C$. We hope that at some point the materials we've found would be investigated experimentally in order to test our theoretical predictions. 

This first application of a high-throughput search of a limited part of the ICSD database, suggests that it possible to find candidates for rare-earth free permanent magnets using a data-mining approach. Further steps in this search would be to release some of the criteria used here, e.g. the limitation of only using 3d and 5d elements, and a maximum of three elements for a suitable compound. Such investigations are underway. We also demonstrated that a post analysis of the output of the high throughput study can be used to further optimize the magnetic properties and identify novel phases. 

\section{Acknowledgement}

Authors would like to acknowledge the support of the Swedish Foundation for Strategic Research, the Swedish Energy Agency, the Swedish Research Council, The Knut and Alice Wallenberg Foundation, eSSENCE, STandUPP and the CSC IT Centre for Science and the Swedish National Infrastructure for Computing (SNIC) for the computation resources. 
 
\bibliography{aipsamp}

\appendix
\section{Materials discarded during the final steps}

\begin{table}[h!]
\centering
\caption{Materials that were considered more thoroughly after the high-throughput step, which however did not fulfill the requirements of a good permanent magnet. NM stands for non-magnetic, NC means non-collinear magnetic ordering}
\begin{tabular}{l l l l l l } 
 \hline \hline
 Material & ICSD & Space & Mag. & MAE (th) & Sat. magn.  \\ 
  & number & group & state &  MJ/m$^3$ & T \\
 \hline
Al$_8$Fe$_4$Hf & 607535 & 139 & FM & -1.4 & 0.28 \\ 
Fe$_7$W$_6$ & 634060  & 166  & FM  & -4.6  & 0.63 \\
CoHfSn & 623801 & 189 & FM & -0.65 & 0.18 \\
Mn$_2$Ta & 109357 & 194 & NC & 1.65 & 0.54  \\
GaMnPt & 103807 & 194 & NC & 3.80 & 0.95 \\
FeHfCl$_6$ & 39817  & 163  & AFM  & {\color{red} } & \\
GeMnTa & 637098  & 189 & AFM  & {\color{red} } & \\
FeTa$_2$B$_2$ & 614207  & 127  & AFM  & {\color{red} } & \\
W$_2$CoB$_2$ & 16776  & 71 & AFM  & {\color{red} } & \\
Co$_2$Ta & 108151 & 194 & AFM  & {\color{red} } & \\
W$_2$MnB$_2$ & 44449  & 127 & NM  & {\color{red} } & \\
W$_2$NiB$_2$ & 615069  & 71 & NM  & {\color{red} } & \\
Ta$_4$FeP & 86378  & 124 & NM  & {\color{red} } & \\
FeGa$_2$Hf$_6$ & 631770  & 189 &  NM & &  \\
Hf$_6$CoBi$_2$ & 54566 & 189 &  NM & &  \\
CoGa$_2$Hf$_6$ & 623077 & 189 & NM & &  \\
CoGaHf & 623085 & 189 & NM & &  \\
 \hline \hline
\end{tabular}
\label{table:4}
\end{table} 

\section{Additional structural information for the  materials found suitable to be used as permanent magnets}

\begin{table}[h!]
\centering
\caption{Additional structural information (ICSD number, space group, crystal system, cell volume, and magnetic moment of the 3d-element) for the three materials that previously have not been investigated for the magnetocrystalline anisotropy (apart from Pt$_2$FeNi that which was studied in the form of L1$_0$ alloy films \cite{FeNiPt}).}
\begin{tabular}{c c c c c c } 
 \hline \hline
 Material & ICSD & Space & Crystal & Cell &  M \\ 
  & number & group & system &  volume,  & of 3d-el., \\
   & & & & $\mathrm{\mathring{A}}^3$ & $\mu_B$ \\
 \hline
Pt$_2$FeNi & 42564 & 123 & tetragonal & 53.83 & 3.21 (Fe)\\ 
& & & & & 0.86 (Ni) \\ 
Pt$_2$FeCu & 53259 & 123 & tetragonal & 54.16 & 3.17 (Fe) \\
& & & & & 0.12 (Cu) \\
W$_2$FeB$_2$ & 43016 & 127 & tetragonal & 102.37 & 1.66 (Fe) \\
 \hline \hline
\end{tabular}
\label{table:5}
\end{table}

\end{document}